\documentstyle[preprint,12pt,aps,epsf]{revtex}
\begin{document}

\title{Charged excitons in doped extended Hubbard model systems.}
\author{J. van den Brink, R. Eder and G.A. Sawatzky}
\address{
Laboratory of Applied and Solid State Physics, Materials Science Centre,\\
University of Groningen,Nijenborgh 4, 9747 AG Groningen, The Netherlands}
\date{\today}
\maketitle

\begin{abstract}

We show that the charge transfer excitons in a Hubbard model system
including nearest neighbor Coulomb interactions effectively attain some
charge in doped systems and become visible in photoelectron and
inverse photoelectron spectroscopies. This shows that the description of
a doped system by an extended Hubbard model differs substantially from
that of a simple Hubbard model. Longer range Coulomb interactions cause
satellites in the one electron removal and addition spectra and the
appearance of spectral weight if the gap of doped systems at energies
corresponding to the excitons of the undoped systems. The spectral
weight of the satellites is proportional to the doping times the coordination
number and therefore is strongly dependent on the dimension. \\
\\
pacs 71.10.-w, 71.35.-y \\
\end{abstract}

The possible importance of nearest neighbor Coulomb interactions in the
description of strongly correlated systems like the 3d transition metal
oxides, low dimensional organic charge transfer salts as well as solid C$_{60}$
has been emphasized in a number of 
studies~\cite{Varma87,Emery79,Bruehwiler93,Stechel95,Janner95,Brink95}.
Although it is clear that a nearest neighbor Coulomb interaction will not
reduce the gap for charged excitations in half filled Hubbard system with
an on site interaction larger than the one electron band width as we
recently showed~\cite{Meinders94} it has also been emphasized by 
Varma~\cite{Varma95}
that such interactions may be very important in describing doped systems
like the high T$_{\rm c}$ superconductors. 
In undoped half filled Hubbard
models nearest neighbor Coulomb interactions 
introduce local charge conserving excitonic states at energies below the
charged excitation conductivity gap which are visible in optical
spectroscopies. These states are not visible in one electron
removal or addition spectroscopies since they do not carry a charge. The
presence of such states is important, however, since they enhance the
nearest neighbor superexchange interactions which involve 
virtual excitations of which the lowest in energy will be these
excitonic states~\cite{Eder95}. The situation is quite different in doped systems.
For example in a hole doped otherwise half filled system there will be
electron addition states corresponding to adding an electron to an already
occupied site with a hole in the nearest neighbor site. Such a state will
be at an energy corresponding to  that of the exciton in the undoped
system. In fact for higher doping there is even the possibility of adding an
electron to an atom which has more than one hole on neighboring sites
providing even lower energy electron addition states because of the
nearest neighbor Coulomb interactions. The spectral weight for these
states in the electron addition spectrum appearing inside the original
Hubbard gap will be stolen from the upper Hubbard band. This effect
will cause the upper Hubbard band to collapse in intensity while moving
up in energy even more rapidly than in a doped Hubbard system. In other
words the spectral weight transfer from the high to the lower energy
scales in the electron addition spectrum will be even more dramatic than
that discussed for a Hubbard system by 
Eskes {\it et al.}~\cite{Eskes91}.

\begin{figure}
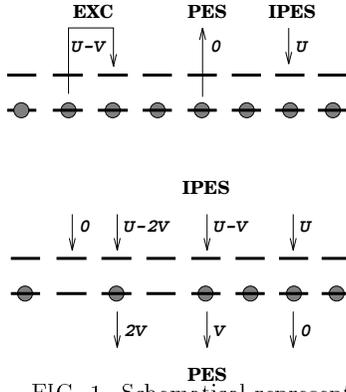

\caption{Schematical representation of a 
extended Hubbard model in the localized 
limit, the upper part for an undoped system,
the lower part for a doped system.
The energies for electron removal (pes),
electron addition (ipes) and (optical) excitation are shown.
A $\bullet$ represents an up or down electron.}
\label{fig:localized}
\end{figure}

The physical picture of what happens in the zero bandwidth limit is
shown in Fig.~\ref{fig:localized}, taking U as the on site energy 
and V as the nearest neighbor electron
repulsion. This picture also clearly shows where the spectral weight in the
gap is coming from. It is interesting to note that for V close to the spin
density wave to charge density wave transition value (V$_{\rm c}$=U/2 in 1d) the
electron addition states corresponding to U-2V will be close to energy
zero and will contribute to the low energy scale physics. This particular
electron addition state corresponds to a state with one hole in an otherwise
half filled Hubbard system accompanied by a nearest neighbor electron
hole excitation and bound to it. This is an oversimplified picture of a
possible 3 particle resonance if V is large enough as envisaged by Varma
{\it et al.}~\cite{Varma87} for the high T$_{\rm c}$'s. 
In this paper we want to give this very simple
picture some theoretical justification with  model calculations including
also a finite band width. We will also look at the influence of finite
temperature and will study the consequences for the photoelectron
spectrum of a system like off stochiometric K$_3$C$_{60}$.

To study this we use the Hamiltonian for the, orbitally non degenerate, 
extended Hubbard model given by:
\begin{eqnarray}
H =&-t& \sum_{\langle ij \rangle, \sigma} (c^\dagger_{i\sigma}c_{j\sigma}
   + h.c.) +  U \sum_i n_{i\uparrow} n_{i\downarrow} \nonumber\\
&+& V \sum_{\stackrel{\langle ij \rangle}{\sigma,\sigma^\prime}}
   n_{i\sigma} n_{j\sigma^\prime},
\label{eq:ham}
\end{eqnarray}
where $n_{i\sigma} \equiv  c^\dagger_{i\sigma}c^{}_{i\sigma}$ and
$c^\dagger_{i\sigma}$ ($c^{}_{i\sigma}$) creates (annihilates) an electron
(or hole) on site $i$ with spin $\sigma = \uparrow$ or $\downarrow$.
The hybridization is denoted by $t$ and a nearest-neighbor pair by 
$\langle ij \rangle$. At half filling the system is in a spin density
wave state for $V < V_c$ and in a charge density wave state for $V > V_c$,
where $V_c \approx U/Z$ and $Z$ is the coordination number of the atoms 
in the lattice~\cite{Hirsch84}.

\begin{figure}
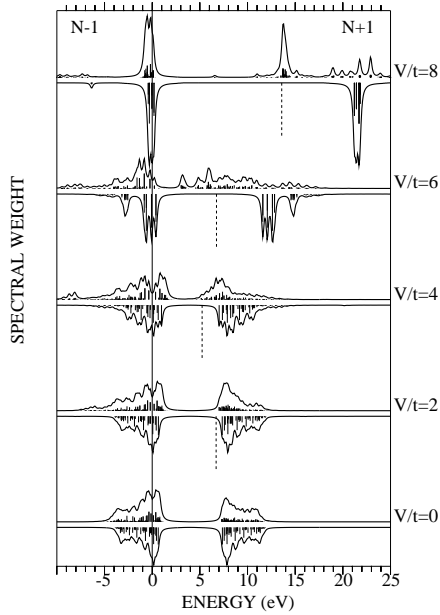

\caption{Electron addition and removal spectrum for a 14 site Extended Hubbard
model at half filling (downward), and with a doping of 2 holes (upward). 
The spectrum for the half filled system is shifted so that the chemical
potential lies in the top of the valence band. The dashed line corresponds
to the onset of the optical spectrum of the half filled system, 
where the zero of energy is taken at the chemical potential. U/t =10.}
\label{fig:spec_w4}
\end{figure}

We performed exact diagonalization calculations of this Hamiltonian for
a one dimensional 14 site system with periodic boundary conditions
looking at both the undoped and the two hole doped system. We take $U$
large compared to the band width i.e. $U=10t$ and vary $V$ from 0 to $8t$
going through the spin density wave -charge density wave transition for
$V=5t$ at half filling. 
We looked at the electron removal, electron addition and the optical
excitations in each case. The results are shown in Fig.~\ref{fig:spec_w4}. 
The electron
addition and removal spectra are displayed towards the positive vertical
direction for the doped system and to the negative direction for
the half filled system. 
The spectrum for the half filled system is shifted in energy so that the
spectral weight below the chemical potential corresponds to the number of 
electrons in the system with two holes.
The dashed line corresponds to the onset of the optical spectrum for
the half filled system, where the zero of enegy is taken at the chemical
potential.
In this figure we see exactly what we have described above
in the simple picture in the atomic limit aside from somewhat broadened
structures shifted in energy and with a large spectral weight
transfer all due to the finite values of the band width. 
The undoped system shows the large correlation gap in the charged
excitations and the appearance of the excitonic states in the gap in the
optical spectrum for non-zero values of $V$. The doped system exhibits a
strong broadening of both the electron removal and addition spectra with
$V$, and shows a large amount of spectral weight appearing in the correlation gap
for finite $V$.
In addition to this filling in of the pseudo gap in the doped system as
V increases we also see the appearance of satellites in the electron
removal part of the spectrum which correspond the removal of an
electron from a neighboring atom to that with a doped hole or from an
atom with two such neighbors. These satellites will be centered at about 
V and
2V below the chemical potential, see Fig.~\ref{fig:localized}. 
The influence of t is to cause a mixing of
these various groups of states causing them to repel each other and
causing a spectral weight transfer towards the lower  energy scale.

\begin{figure}
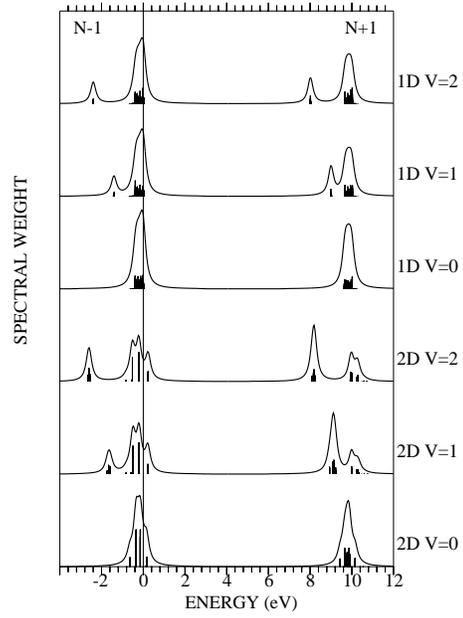

\caption{Electron addition and removal spectrum small bandwidth
limit for ten sites in one and two dimensions.
The doping is 0.1, U= 10eV, and t= 0.1eV}
\label{fig:spec_t01}
\end{figure}

It is interesting to look at the spectral weights in more detail since they
will be strongly influenced by the dimensionality of the system or rather
the coordination number. We consider first doping with only one hole.
In the atomic limit and for V=0 there are 3 poles in the
(inverse)photoemission spectrum, one below the chemical potential with
a weight of N(1-x), one just at the chemical potential with a weight 2x
and one at U above above the chemical potential also with a weight of N(1-x) 
where N is the number of sites and x the hole doping (discussed in
detail~\cite{Eskes91}).
The pseudo gap in the electron addition spectrum
as discussed above is U. A small band width  (finite t) broadens these
states into bands and causes spectral weight to be transfered from the
states at $U$ to the chemical potential in the electron addition part of the
spectrum. For finite V the situation changes markedly. Now there are in
total 5 poles: one at -V with a weight of Zx, one just below the chemical
potential with a weight of N(1-x)-Zx, one just above the chemical
potential with a weight 2x one at U-V with a weight Zx and at U with a
weight N(1+x)-Zx.
The spectrum for ten sites is shown in Fig.~\ref{fig:spec_t01}. 
The spectral weight of the
satellites  in this limit of small doping and small t scales with the
coordination number  so that their intensities will be two times as large in
2D than the 1D spectra shown in Fig.~\ref{fig:spec_w4}.  
At higher doping
concentrations also more poles can appear,in the atomic limit at energies
of -nV, nV and U-nV where n is an integer between 1 and Z in analogy to
the discussion above for the two hole doped 1 D system.  Here n
corresponds to the number of occupied and empty neighboring sites in the
final state. We see therefore that the pseudo gaps in the spectrum occuring
for small V of magnitude U will quickly fill up with spectral weight and
the total energy spread of the spectrum will increase rapidly as V and the
doping are increased. 

In this atomic limit the intensities of the various
peaks will be determined solely by the statistical distribution of possible
configurations. This will change as we switch on the band width due to
t. Increasing t will of course further increase the spread of the spectrum
however as demonstrated in the V=0 case~\cite{Eskes91} spectral weight will
rapidly be transfered towards the lower energy scale from the higher
energy scales untill finally for large enough t all signs of the higher energy
features will disappear. The final result at large t will be a band of states
concentrated around the chemical potential with a total width of about
2Zt as in one electron theory. In this limit one electron band theory will
be valid.

\begin{figure}
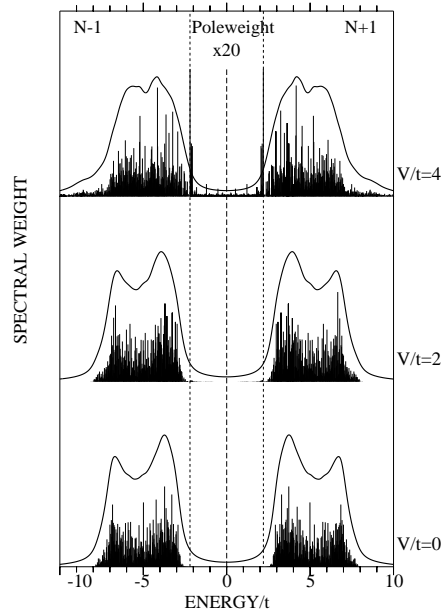

\caption{Electron addition and removal spectrum at a temperature
$kT/t=1$ with $U/t=10$ for $V/t=0$,2 and 4, for a half filled 8-site
Extended Hubbard ring. The intensities of the poles between the dashed
lines are amplified by a factor of 20.}
\label{fig:temp}
\end{figure}

At finite temperature the  spectral function is given by:
\begin{eqnarray}
A(k,\omega) =  \frac{1}{Z}&& \ \sum_{a,b,\sigma} \delta(\omega-E_b+E_a)
	     \ \langle a | c_{k \sigma} | b \rangle \nonumber \\ 
             &&\langle b | c^{\dagger}_{k \sigma} | a \rangle
             \ e^{-\beta E_a}.
\label{eq:sp_temp}
\end{eqnarray}
where $| b \rangle$ represents the final states, and $\beta = 1/kT$.
The temperature dependence is determined by the Boltzmann-factor 
$e^{-\beta E_a}$ for the initial state $| a \rangle$
and the partition function is given by $Z= Tr\ e^{-\beta H}$.
Note that the sum in~(\ref{eq:sp_temp}) involves all final and all 
initial states, so that for an exact diagonalization of the 
Hamiltonian~(\ref{eq:ham}) we need all eigenvalues and eigenstates of both 
initial and final states, which limits our cluster sizes considerably.
In Fig.~\ref{fig:temp} the spectral function at half filling as a function 
of temperature
for U/t=10, and with V/t=0,2 and 4, for a one dimensional eight 
site cluster is shown.
As the temperature increases, states inside the gap in the EHM gradually
obtain more weight, as opposed to the simple Hubbard model, where there
is no weight inside the gap observed. 
This is easy to understand
if one realizes that the effect of the temperature is to occupy excited states
of the N-particle system, which in the case of the extended Hubbard model, 
include the states with a 
nearest neighbor exciton. If an electron is removed from such
an excited state, which is closer to the vacuum level, the energy of
this electron is lowered with the exciton binding energy.
So the situation at finite temperature is similar to that at finite doping,
be it that at finite temperature the exciton is present in the initial state,
and at finite doping, the exciton is present in the final state.
The gradual filling in of the gap in the EHM is directly due to the 
correlations present in this model. For instance in a band semiconductor 
described by one electron theory, 
the gap remains intact even at elevated temperature, but some weight occurs
in photo emission corresponding to the termally excited electrons in
the conduction band.

\begin{figure}
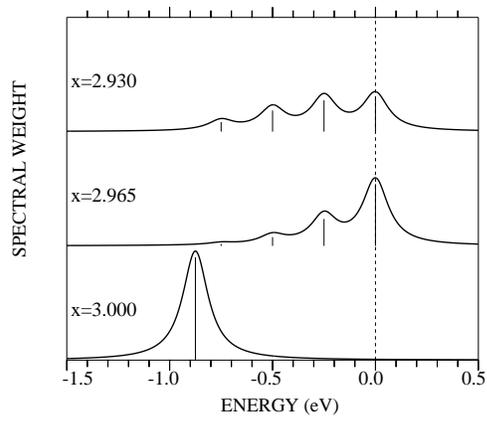

\caption{Electron removal spectrum for 3-fold degenerate EHM on a FCC
lattice in the zero 
bandwidth limit for x=3-$\delta$ electrons per site. $U=1.75$ eV and
$V=0.25$ eV. In the x=3 spectrum we took the chemical potential to
be in the middle of the gap.}
\label{fig:bino}
\end{figure}

Let us apply these observations to C$_{60}$ and its potassium doped 
derivatives, which are described by an
orbital degenerate extended Hubbard model.
Photoemission experiments on K$_3$C$_{60}$, 
show that this compound is metallic but that the low energy part
in the spectrum is broadened over a range of about
1 eV~\cite{Chen91}. 
This feature cannot be a band structure effect since the width of the $t_{1u}$
band is expected to be at most about 0.6 eV~\cite{Louie93}.
and that of the occupied part only 0.3 eV. One possible explanation
involves the smearing out of spectral weight due to vibrational excitations
and plasmons  which can accompany the removal of an electron as argued
by Gunnarsson {\it et al.} and 
Knupfer {\it et al.} ~\cite{Knupfer93,Gunnarsson96}.
An alternative explanation is that the material is 
a slightly off stochiometric strongly correlated system
and therefore a (bad) metal. In this case one would expect
as argued above a spreading out of spectral weight because of satellites
associated with the Coulomb interactions U and V as suggested by 
us~\cite{MeindersPhD}.
The problem with this has been in the
past that rather high non-stochiometries are required to get sufficient
intensity out ot high energies in the model where only U is considered.
However things may improve if we include the now known rather large
values of V~\cite{Bruehwiler93} in addition to U. 
As discussed above the intensities of the
satellites can be rather large as the coordination number Z is 12 for an
FCC lattice. To check this we show  
n Fig.~\ref{fig:bino} the photoemission spectrum for a 3 fold 
degenerate Hubbard
model with zero bandwidth with on average 3-$\delta$ electrons per site is
shown. Even a very small deviation from the x=3 insulating regime shifts
the chemical potential into the lower Hubbard band and gives
low energy satellites with high intensities. 
This may provide an explanation for the large spread in the
photoemission spectra of K$_3$C$_{60}$
in spite of the fact that in K$_3$C$_{60}$ the deviation
from perfect stochiometry is only 0.09~\cite{Fischer95}.
In the case of a finite
bandwidth, we expect, as observed in the 1D calculations above, that the
satellites will shift somewhat towards the chemical potential and that 
weight will be transferred to the low energy part of the spectrum. 

We conclude that the features of a Hubbard model with inter-site Coulomb 
interactions upon doping or at finite temperature are both qualitatively and
quantitatively different from the simple Hubbard model.
In the excitation spectrum at finite doping, the nearest neighbor Coulomb
interaction introduces satellites both in the high and low energy part
of the spectrum and renormalizes the pseudo gap.
It increases the low energy spectral weight. At finite temperatures
states inside the gap obtain spectral
weight, and the gap seems to fill in gradually. 
 We have demonstrated that excitonic states in the
insulating undoped system visible only in optical spectroscopies attain
an effective charge in the doped systems and become visible in the
electron removal and addition spectra. These states quickly fill in the gap
with spectral weight especially in higher dimensions. We have also argued
that the inclusion of the nearest neighbor Coulomb interactions in the case
of C$_{60}$ may provide an explanation for the large spread in the
photoemission spectra.

\end{document}